\documentclass[preprint,aps,prd,groupedaddress,showpacs,nofootinbib]{revtex4-1}%
\usepackage{graphicx}
\usepackage{amsmath}
\usepackage{amssymb}
\usepackage{bm}
\usepackage{epsfig}
\usepackage{dcolumn}
\usepackage{slashed}
\usepackage{empheq}
\usepackage{xcolor}

\def\mpsi{m_{J/\psi}}
\def\mphi{m_{\phi}}

\begin{document}

\title{Rare exclusive decays of the $Z$-boson revisited}

\author{Ting-Chung Huang}
\email{tingchunghuang2014@u.northwestern.edu}
\affiliation{Department of Physics \& Astronomy,Northwestern University, Evanston, IL 60201,USA}

\author{Frank Petriello}
\email{f-petriello@northwestern.edu}
\affiliation{Department of Physics \& Astronomy,Northwestern University, Evanston, IL 60201,USA}
\affiliation{High Energy Physics Division, Argonne National Laboratory, Argonne, IL 60439,USA}

\date{\today}

\begin{abstract}

The realization that first- and second-generation Yukawa couplings can be probed by decays of the Higgs boson to a meson in association with a photon has renewed interest in such rare exclusive decays.  We present here a detailed study of the rare $Z$-boson processes $Z \to J/\psi+\gamma$, $Z \to \Upsilon+\gamma$, and $Z \to \phi+\gamma$ that can serve as benchmarks for the analogous Higgs-boson decays.  We include both direct-production and fragmentation contributions to these decays, and consider the leading QCD corrections and the relativistic corrections to the $J/\psi$ and $\Upsilon$ processes.  We present numerical predictions for the branching ratios that include a careful accounting of the theoretical uncertainties.

\end{abstract}

\maketitle

\section{Introduction}
A primary goal of Run II of the LHC will be the further investigation of the Higgs boson discovered in 2012.  Current measurements by the ATLAS and CMS collaborations indicate that the couplings of this new state agree with Standard Model (SM) predictions at the 20--30\% level~\cite{atlascoup,cmscoup}.  These measurements so far only provide information about the Higgs couplings to electroweak gauge bosons and to third-generation fermions.  The Higgs Yukawa couplings to first- and second-generation quarks are currently unknown.  It is extremely difficult to experimentally access these couplings.  They are predicted to be small in the SM, and the inclusive decays of the Higgs to these states are swamped by large QCD backgrounds.  These couplings are  indirectly constrained weakly by the inclusive Higgs production cross section~\cite{Kagan:2014ila}.  Such constraints only probe the simultaneous deviation of all Yukawa couplings.  They do not allow the separate Yukawa couplings of the various quarks to be determined.

It was discovered recently that it is possible to explore these couplings using rare exclusive decays of the Higgs boson to mesons in association with a photon.  The first manifestation of this idea was the suggestion that the Higgs coupling to charm quarks could be probed using the exclusive decay $H \to J/\psi + \gamma$~\cite{Bodwin:2013gca}.    The enhancement of the branching ratio for this mode compared to initial expectations came from the realization that two distinct production mechanisms give rise to this process:
\begin{itemize}

\item the {\it direct} contribution in which the Higgs boson decays into a $c\bar{c}$ pair, one of which radiates a photon before forming a $J/\psi$;

\item the {\it indirect} or {\it fragmentation} contribution, in which the Higgs boson decays to a $\gamma$ and an off-shell $\gamma^*$, with the $\gamma^*$ then fragmenting into a $J/\psi$. 

\end{itemize}
Initial considerations of this process~\cite{Keung:1983ac} studied only the direct production mechanism.  The indirect production amplitude is larger, and its interference with the direct mode renders this decay measurable at the LHC and sensitive to the $Hc\bar{c}$ coupling.  Although this coupling can possibly be accessed at the LHC using charm tagging~\cite{Delaunay:2013pja}, its phase can only be studied using processes such as this rare decay that involve quantum interference effects.  It was later realized that other exclusive decays of the Higgs boson to light mesons in association with either a photon or a heavy electroweak gauge boson can similarly be used to probe the Yukawa couplings of the other first- and second-generation quarks~\cite{Kagan:2014ila}.  Decays to light mesons together with a heavy gauge boson may also be used to probe the structure of the Higgs couplings to electroweak gauge bosons~\cite{Isidori:2013cla}.  The study of these rare decays at the high-luminosity LHC offers potentially the only way to directly study these couplings in the foreseeable future.  Their predicted rates at planned future $e^+e^-$ machines are too small.  Only the LHC and future very high-energy hadron colliders produce enough Higgs bosons to allow observation of these decays. 

Initial experimental studies of these channels have begun and appear promising.  One topic needed to further assist these investigations is a set of experimental benchmarks besides the Higgs decays that can be used to refine and validate search techniques. Obvious candidates are rare decays of the $Z$-boson. Its mass is not too much smaller than the Higgs mass, and it is also produced primarily at the LHC as an $s$-channel resonance.  The set of rare Higgs boson decays outlined in Refs.~\cite{Bodwin:2013gca,Kagan:2014ila} can be divided into two broad categories based on their experimental signatures:
\begin{itemize}

\item the decays $H \to V+\gamma$ where $V=J/\psi$ or $\Upsilon$ feature the final state $l^+l^- \gamma$ after leptonic decays of the vector quarkonium are required;

\item decays of the Higgs boson to a light meson such as the $\phi$ plus a photon.  In this case the $\phi$ decays hadronically, and a track-based trigger must be developed.

\end{itemize} 
We focus in this manuscript on the decays $Z \to J/\psi +\gamma$, $Z \to \Upsilon+\gamma$ and $Z \to \phi +\gamma$, which are representative of these two categories.  The decays of the $Z$-boson to the heavy quarkonium states $J/\psi$ and $\Upsilon$ were studied in a classic paper by Guberina {\it et al.}~\cite{Guberina:1980dc} (GKPR).  In their work GKPR include only the direct production mechanism, and work in the exact non-relativistic limit.  As far as we are aware the decay $Z \to \phi +\gamma$ has not been studied in the literature.

Our goal in this manuscript is to provide up-to-date theoretical predictions for these rare $Z$-boson decays for use in LHC searches.  We consider both the indirect and direct contributions to both decays.  For the $J/\psi$ and $\Upsilon$ final states we use the non-relativistic QCD (NRQCD) framework~\cite{Bodwin:1994jh} to perform the calculation.  We cross-check our result using the light-cone distribution amplitude (LCDA) approach~\cite{Lepage:1980fj,Chernyak:1983ej}.  The evaluation of the direct amplitude using both approaches allows us to include the leading ${\cal O}(\alpha_s)$ QCD corrections and the leading ${\cal O}(v^2)$ relativistic corrections to the decay.  We compute the $Z \to \phi +\gamma$ decay using the LCDA approach, and include the leading-logarithmic QCD corrections that come from the evolution of the LCDA from the hard scale $M_Z$ down to the phi mass scale, $\mphi$.  We perform a detailed estimate of the remaining sources of theoretical uncertainty affecting both decays.  We find the following final results for the branching ratios:
\begin{empheq}[box=\fbox]{align}
B_{SM}(Z \to J/\psi+\gamma) &= (9.96 \pm 1.86) \times 10^{-8}, \nonumber \\
B_{SM}(Z \to \Upsilon(1S)+\gamma) &= (4.93 \pm 0.51) \times 10^{-8}, \nonumber \\
B_{SM}(Z \to \phi+\gamma) &= (1.17 \pm 0.08) \times 10^{-8}.
\end{empheq}
We define the branching ratios as $B_{SM}(Z\to M+\gamma)=\Gamma(Z\to M+\gamma)/\Gamma(Z)$, where $\Gamma(Z)$ is the total width of the $Z$-boson.  We use $\Gamma(Z)=2.4952$\, GeV from the Particle Data Group~\cite{Agashe:2014kda}.

Although small, it is possible that the heavy quarkonium branching ratios will be accessible in Run II measurements~\cite{expcom}.  Compared to the analogous Higgs-boson decays~\cite{Bodwin:2013gca,Kagan:2014ila}, the $J/\psi$ and $\phi$ branching ratios are smaller by 1-2 orders of magnitude.  This is due primarily to the suppression of the indirect amplitude in the $Z$-boson decays as compared to the Higgs decays.  This amplitude proceeds through the $Z\gamma\gamma^{*}$ effective coupling, which receives contributions from Standard Model anomaly diagrams.  It was previously suggested in the literature that the indirect amplitude could give large contributions to the similar process of a $Z$-boson decaying to a pseudoscalar meson and a photon~\cite{Keum:1993eb}.  We show here that there is no such enhancement for this process.  The indirect amplitude depends on the difference between fermion masses within a generation, and goes to zero for heavy fermions such as the top quark.  The only numerically-relevant contributions therefore come from the tau lepton, the charm quark and the bottom quark.  Since these fermion masses are small, the indirect amplitude is small for this process.  Furthermore, the Landau-Yang theorem~\cite{LY} prevents the decay of the $Z$-boson to two on-shell photons, and therefore requires that the indirect amplitude for the process considered here vanishes in the limit $\mpsi \to 0$.  This implies that there can be no enhancement with respect to the direct amplitude by the ratio $m_H^2/\mpsi^2$, as there is for the analogous Higgs decays.  These effects leads to an indirect amplitude with a magnitude less than 1\% of the direct-amplitude magnitude.

Our paper is organized as follows. In Section~\ref{sec:jpsi}, we derive the amplitude for the $Z \to J/\psi +\gamma$ decay.  The $\Upsilon$ decay calculation is identical.  We discuss our evaluation of both the direct and indirect contributions, and our evaluation of the leading QCD and relativistic corrections.  In Section~\ref{sec:phi} we describe our calculation of the $Z \to \phi +\gamma$ process using the LCDA approach.  We present our numerical results and describe our estimates of the theoretical uncertainties in Section~\ref{sec:numerics}.  We conclude in Section~\ref{sec:conc}.

\section{The decay $Z\rightarrow J/\psi + \gamma$}
\label{sec:jpsi}

\begin{figure}[!h]
\includegraphics[scale=0.35]{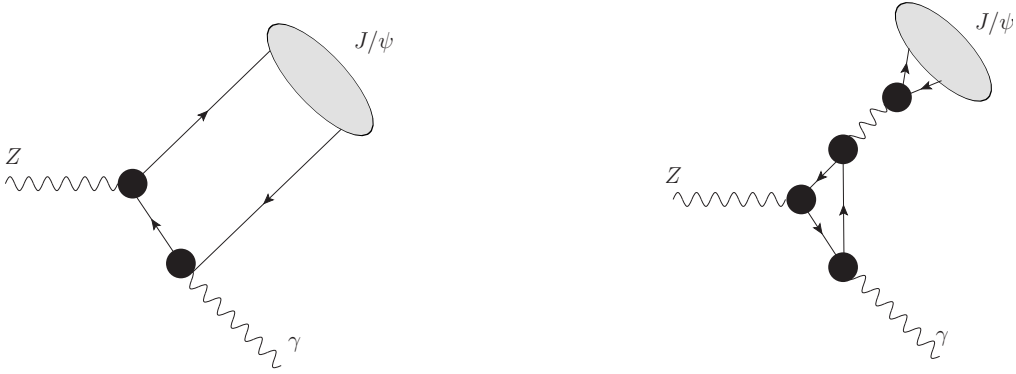}
\caption{\label{fig:diagrams} Representative Feynman diagrams contributing to the direct amplitude (left panel) and indirect amplitude (right panel) for $Z\rightarrow J/\psi + \gamma$.  Similar diagrams lead to the process $Z\rightarrow \phi + \gamma$ and $Z\rightarrow \Upsilon + \gamma$.  We note that the direct amplitude receives contributions from two diagrams.  The second diagram is obtained by reversing the fermion flow in the diagram shown here. }
\end{figure}

We begin by discussing the decay $Z\rightarrow J/\psi + \gamma$.  Since the calculation of the $\Upsilon$ decay is identical to the $J/\psi$ we do not present it explicitly.  We give numerical results for both modes in a later section.  This process receives contributions from both a direct amplitude and an indirect amplitude.  These are shown respectively in the left and right panels of Fig.~\ref{fig:diagrams}.  We calculate the direct-amplitude contribution to this process using the non-relativistic QCD (NRQCD) framework~\cite{Bodwin:1994jh}.  We include the velocity corrections through ${\cal O}(v^2)$.  In addition, we include the leading ${\cal O}(\alpha_s)$ corrections using the light-cone distribution amplitude (LCDA) approach~\cite{Lepage:1980fj,Chernyak:1983ej}.  The indirect amplitude proceeds through the loop-induced $Z\gamma^{*}\gamma$ effective vertex, which can be calculated in perturbation theory.  The subsequent $\gamma^{*} \to J/\psi$ transition can be obtained from data.

We perform our calculation to leading-order in the ratio $\mpsi^2/M_Z^2$. The corrections from the higher-order terms in this expansion are expected to be at the $0.1\%$ level, far below any other source of theoretical error we consider.  We have checked that a certain class of these corrections which we can easily obtain (those coming from the final-state phase space and from the direct amplitude) have no effect on our numerical results.

\subsection{The direct amplitude in the non-relativistic limit}

We begin by calculating the direct amplitude in the non-relativistic $v=0$ limit.  We have reproduced and have found agreement with the result in GKPR~\cite{Guberina:1980dc}.  We briefly sketch the derivation here.

We define the partonic process leading to $J/\psi$ production as
\begin{equation}
Z(P) \rightarrow c(p_{1})\bar{c}(p_{2})+\gamma(p_{\gamma}).
\end{equation}
We introduce the relative momenta between the $c$ and $\bar{c}$ as $q = (p_1-p_2)/2$, and the total momentum of the $J/\psi$ as $p_{V}= 2p = p_1+p_2$.  We then have the following relations among the momenta:
\begin{eqnarray}
p_{1} &=& p + q, \;\;\; p_{2}= p - q,  \;\;\; p \cdot q = 0,\nonumber \\
p^{2}_{1} &=& p^{2}_{2}=m^{2}_{c}, \;\;\; p^2 = E^2,\;\;\; q^2 = m_c^2-E^2=-m_c^2 v^2.
\label{eq:momdefs}
\end{eqnarray}
In order to produce a $J/\psi$ the $c\bar{c}$ pair must be produced in a spin-triplet, color-singlet final state. We use a projection operator~\cite{Guberina:1980dc,Bodwin:2002hg} to enforce the production of this final state:
\begin{equation}
\label{eq:trace}
{i\cal M}_{direct}=\sqrt{2m_{J}}\phi_{0}(J)\text{Tr}[(i{\cal M}_{c\bar{c}\gamma})\Pi_{1}(p,q,\epsilon^{*})],
\end{equation}
where the projector is given by
\begin{equation}
\Pi_{1}(p,q,\epsilon^{*})=\frac{1}{8\sqrt{2}E^2(E+m_{c})}(\slashed{p}_{2}-m_{c})\slashed{\epsilon}^{*}
(\slashed{p}_{1}+\slashed{p}_{2}+2E)(\slashed{p}_{1}+m_{c}) \otimes \frac{{\bf 1}}{\sqrt{N_c}}.
\label{eq:proj}
\end{equation}
The amplitude ${\cal M}_{c\bar{c}\gamma}$ is obtained by directly calculating the Feynman diagrams from the left panel of Fig.~\ref{fig:diagrams} in QCD perturbation theory.  Summing the two diagrams which contribute to the direct amplitude yields
\begin{eqnarray}
i{\cal M}_{c\bar{c}\gamma} &=&
(ig_{Z}\slashed{\varepsilon}_{Z}(g^{c}_{V}-g^{c}_{A}\gamma_{5}))
\frac{i(-\slashed{p}_{2}-\slashed{p}_{\gamma}+m_{c})}{(p_{2}+p_{\gamma})^{2}-m^{2}_{c}}(-ieQ_{c})\slashed{\varepsilon}^{*}_{\gamma}\nonumber \\
&+& (-ieQ_{c})\slashed{\varepsilon}^{*}_{\gamma}
\frac{i(\slashed{p}_{1}+\slashed{p}_{\gamma}+m_{c})}{(p_{1}+p_{\gamma})^{2}-m^{2}_{c}}
(ig_{Z}\slashed{\varepsilon}_{Z}(g^{c}_{V}-g^{c}_{Z}\gamma_{5})).
\end{eqnarray}
The external spinors associated with the quark and anti-quark appearing in the partonic amplitude have been removed from this expression; they are replaced by the projector of Eq.~(\ref{eq:proj}) when performing the trace indicated in Eq.~(\ref{eq:trace}). We have included the NRQCD matrix element to convert this into the $J/\psi$ amplitude in Eq.~(\ref{eq:trace}), resulting in the appearance of $\phi_{0}(J)$, the $J/\psi$ wave-function at the origin.  The trace is over both the Dirac and color indices.

To obtain the non-relativistic expression, we set $q=0$ in Eq.~(\ref{eq:proj}).  We note that this also sets $v=0$ upon using the kinematic relations in Eq.~(\ref{eq:momdefs}).  Since $\mpsi=2m_c \sqrt{1+v^2}$, this also has the affect of enforcing $\mpsi=2m_c$ in the non-relativistic limit.  After a straightforward calculation using these relations we arrive at the result
\begin{equation}
{\cal M}^{(0)}_{J/\psi,direct}=i\frac{8\sqrt{N_{c}}eQ_{c}g_{Z}g_{A}^c \phi_{0}(J)\sqrt{\mpsi}}{M^{2}_{Z}}
\epsilon_{\alpha\mu\nu\rho}\varepsilon^{\alpha}_{Z}\varepsilon^{*\mu}_{\gamma}\varepsilon^{*\nu}_{J/\psi}
p^{\rho}_{\gamma}.
\label{eq:nonrel}
\end{equation}
The superscript denotes that neither relativistic corrections nor higher-order perturbative QCD corrections have been included.  As noted earlier, we have kept only the leading term in the $\mpsi^2/m_Z^2$ expansion.   For the electromagnetic coupling $e$ we use the value at zero momentum transfer, as appropriate for an on-shell photon.  $Q_c=2/3$ is the charm-quark charge.  $g_Z$ denotes the overall coupling strength of the Z-boson to fermions, while $g_{V,A}^f$ denote the vector and axial couplings of the fermion $f$.  We write these as follows:
\begin{equation}
g_{Z} = 2 \times 2^{1/4}\sqrt{G_{F}} M_{Z}, \;\;\; g_V^f = \frac{I_3^f}{2}-Q_f \, \sin^{2}\theta_{W},\;\;\; g_{A} = \frac{I_3^f}{2},
\end{equation}
where $G_F$ is the Fermi constant, $\sin \theta_W$ is the sine of the weak-mixing angle, and $I_3^f=\pm 1/2$ for up-type and down-type quarks, respectively.  We note that the amplitude is proportional to the axial coupling of the charm quark. If this quantity was zero, Furry's theorem would require that this amplitude was vanishing.

\subsection{Relativistic corrections in NRQCD}

To obtain the leading relativistic corrections to this amplitude, we follow the approach outlined in Ref.~\cite{Bodwin:1407}.  We keep the $q$-dependence in the projector of Eq.~(\ref{eq:proj}), and expand the result in the parameter $v^2$ introduced in Eq.~(\ref{eq:momdefs}).  We keep only the ${\cal O}(v^2)$ correction.  This correction is a ratio of NRQCD matrix elements:
\begin{equation}
\label{eq:convertNRQCD}
v^{2n} \to \langle v^{2n}\rangle=
\frac{1}{m_Q^{2n}} 
\frac{\langle V(\bm{\epsilon})
|\psi^\dagger(-\frac{i}{2} 
\tensor{\bm{\nabla}})^{2n}\bm{\sigma}\cdot\bm{\epsilon}\chi|0\rangle}
{\langle V(\bm{\epsilon})
|\psi^\dagger \bm{\sigma}\cdot\bm{\epsilon}\chi|0\rangle}.
\end{equation}

We being by evaluating the trace over the projection operator and partonic amplitude in Eq.~(\ref{eq:trace}), and averaging over the spatial direction of the momentum $q$ in the $J/\psi$ rest frame using the following operation:
\begin{equation}
{\int}_{\!\!\!\!\hat{\bm{q}}} 
\equiv\int\!\frac{d \Omega_{\hat{\bm{q}}}}{4 \pi},
\end{equation} 
%
where we have defined $\hat{\bm{q}} = \bm{q}/|\bm{q}|$.  Doing so, we find the following four auxiliary integrals over $\hat{\bm{q}}$ that are needed for this calculation:
\begin{eqnarray}
I &=& {\int}_{\!\!\!\!\hat{\bm{q}}}\,
\frac{p\cdot p_\gamma}{(p-q)\cdot p_\gamma},
\\
I^\mu &=&  {\int}_{\!\!\!\!\hat{\bm{q}}}\,
\frac{p\cdot p_\gamma}{(p-q)\cdot p_\gamma}\,q^\mu,
\\
I^{\mu \nu} &=&  {\int}_{\!\!\!\!\hat{\bm{q}}}\,
\frac{p\cdot p_\gamma}{(p-q)\cdot p_\gamma}\,q^\mu q^\nu.
\end{eqnarray}
The analytic expressions for these integrals have been derived in Ref.~\cite{Bodwin:1407}, leading to the results
\begin{eqnarray}
I_0 &=&  =
\frac{1}{2 \delta} \log \frac{1+\delta}{1-\delta}, \nonumber \\
I^\mu &=& \frac{4E^2(1-I_0)}{m_H^2 - 4E^2} \,\bar p^\mu_\gamma \equiv I_1\,\bar p^\mu_\gamma, \nonumber \\
I^{\mu \nu}&=& 
\frac{E^2-m_Q^2 I_0}{2}
\left(-g^{\mu \nu}+\frac{p^\mu p^\nu}{p^2}\right)
+\frac{8E^2 [ (m_Q^2 + 2E^2)I_0 - 3 E^2]}{(m_H^2 -4E^2)^2} 
\,\bar p^\mu_\gamma \bar p^\nu_\gamma \nonumber \\
&\equiv & I_{2a} \left(-g^{\mu \nu}+\frac{p^\mu p^\nu}{p^2}\right) + I_{2b} \,\bar p^\mu_\gamma \bar p^\nu_\gamma .
\end{eqnarray}
where we have introduced the abbreviations
\begin{equation}
\delta=\frac{v}{\sqrt{1+v^2}},\;\;\; \bar p_\gamma = p_\gamma - \frac{p_\gamma \cdot p}{p^2} p.
\end{equation}
We note that the kinematic relations in Eq.~(\ref{eq:momdefs}) have been used in arriving at the expression for $\delta$.  Since we are expanding our amplitude in both $v^2$ and $m_c^2/M_Z^2$, we need to obtain only the leading terms in these small quantities.  We find
\begin{eqnarray}
\label{eq:auxints}
I_0 &=& 1+\frac{v^2}{3} +{\cal O}(v^4),\;\;\; I_1 = -\frac{4}{3} v^2 \frac{m_c^2}{M_Z^2}+{\cal O}(v^4,m_c^4/M_Z^4), \nonumber \\
I_{2a} &=& \frac{1}{3} v^2 m_c^2+{\cal O}(v^4),\;\;\; I_{2b} = {\cal O}(v^4,m_c^4/M_Z^4).
\end{eqnarray}
We arrive at the following result for the direct amplitude of Eq.~(\ref{eq:trace}) expanded to ${\cal O}(v^2)$ in terms of these quantities:
\begin{equation}
\begin{split}
{\cal M}^{v^2}_{J/\psi,direct}& =i\frac{8\sqrt{N_{c}}eQ_{c}g_{Z}g_{A}^c\phi_{0}(J)\sqrt{\mpsi}}{M^{2}_{Z}}
\epsilon_{\alpha\mu\nu\rho}\varepsilon^{\alpha}_{Z}\varepsilon^{*\mu}_{\gamma}\varepsilon^{*\nu}_{J/\psi}
p^{\rho}_{\gamma} \left[ I_0 \left(1-\frac{3v^2}{8} \right) +\frac{5}{8} \frac{I_{2a}}{m_c^2} \right] \\ &+{\cal O}(v^4,m_c^2/M_Z^2).
\end{split}
\end{equation}
We next use the replacement of Eq.~(\ref{eq:convertNRQCD}) to convert the $v^2$ appearing in this result into a ratio of NRQCD matrix elements, whose numerical values are known.  Incorporating this replacement and the integrals of Eq.~(\ref{eq:auxints}) yields our final result:
\begin{equation}
\label{eq:ampdirrel}
{\cal M}^{v^2}_{J/\psi,direct}=i\frac{8\sqrt{N_{c}}eQ_{c}g_{Z}g_{A}^c\phi_{0}(J)\sqrt{\mpsi}}{M^{2}_{Z}}
\epsilon_{\alpha\mu\nu\rho}\varepsilon^{\alpha}_{Z}\varepsilon^{*\mu}_{\gamma}\varepsilon^{*\nu}_{J/\psi}
p^{\rho}_{\gamma}\left(1+\frac{\langle v^{2} \rangle}{6}\right).
\end{equation}
Although we will discuss numerics in more detail later in our draft, we note that $ \langle v^{2}\rangle = 0.20$, leading to a $3\%$ increase in the direct amplitude from relativistic corrections.

\subsection{The LCDA approach to the $Z\rightarrow J/\psi + \gamma$ decay}
\label{sec:LCDA}

We note that since $M_Z \gg \mpsi$, this process consists of a photon recoiling against an energetic $J/\psi$, with both the $c$ and $\bar{c}$ moving along the light-cone direction defined by the $J/\psi$ momentum.  This picture allows us to use LCDA techniques~\cite{Lepage:1980fj,Chernyak:1983ej} to calculate the direct amplitude to leading order in $\mpsi^2/M_Z^2$.  The advantage of this approach is that the leading ${\cal O}(\alpha_s)$ QCD corrections are known in the LCDA approach, and can be used to improve our prediction.  Furthermore, the LCDAs satisfy an evolution equation that can be used to sum the leading-logarithmic corrections arising from collinear gluon emission.  Since we find that the QCD corrections are small, we do not included this renormalization-group improvement in our result.

We begin by introducing the following light-cone momentum directions:
\begin{equation}
n^{\mu} = (1,0,0,1),\;\;\; \bar{n}^{\mu} = (1,0,0,-1).
\end{equation}
We align $n$ to lie along the $J/\psi$ direction, and $\bar{n}$ to lie along the photon direction.  We can express all momenta in terms of these directions:
\begin{eqnarray}
\label{eq:LCDAmom}
p^{\mu}_{\gamma} &=& \frac{M^{2}_{Z}-\mpsi^2}{2M_{Z}} \bar{n}^{\mu},\;\;\; p^{\mu}_{V} = 
	\frac{M_{Z}}{2}n^{\mu}+\frac{\mpsi^2}{2M_{Z}}\bar{n}^{\mu}, \\
p^{\mu}_{1} &=& u\frac{M_{Z}}{2}n^{\mu}+\frac{m^2_{c}}{2u M_{Z}}\bar{n}^{\mu},\;\;\;
p^{\mu}_{2} = \bar{u}\frac{M_{Z}}{2}n^{\mu}+\frac{m^2_{c}}{2\bar{u}M_{Z}}\bar{n}^{\mu}.
\end{eqnarray}
We have introduced the light-cone momentum fractions $u$ and $\bar{u}$ that parameterize the fractions of the $J/\psi$ light-cone momentum carried by the $c$ and $\bar{c}$, respectively.  We note that $\bar{u}=1-u$.  These quantities satisfy the relation $\mpsi^2=m_c^2/(u\bar{u} )$.  

To proceed with the LCDA approach, we expand these kinematic relations to leading order in the ratios $\mpsi^2/M_Z^2$ and $m_c^2/M_Z^2$, leading to the simplified expressions
\begin{equation}
p^{\mu}_{\gamma} \approx \frac{M_{Z}}{2} \bar{n}^{\mu},\;\;\; p^{\mu}_{V} \approx 
	\frac{M_{Z}}{2}n^{\mu},\;\;\; p^{\mu}_{1} \approx u\frac{M_{Z}}{2}n^{\mu},\;\;\;
p^{\mu}_{2} \approx \bar{u}\frac{M_{Z}}{2}n^{\mu}.
\end{equation}
The momentum fraction $u$ takes on values in the range $[0,1]$.  We then compute the diagrams corresponding to the direct amplitude in Fig.~\ref{fig:diagrams} using standard Feynman rules.  We then use projection operators~\cite{Beneke:2001} to extract the amplitude for $J/\psi$ production in terms of the appropriate LCDAs.  There are two relevant projection operators to consider: one which describes the production of a transversely-polarized $J/\psi$, and one which describes the production of a longitudinally polarized $J/\psi$.  We find that the production of a transversely-polarized $J/\psi$ vanishes to leading order in $\mpsi^2/M_Z^2$.  The production rate of a longitudinally polarized $J/\psi$ is not similarly suppressed.  This agrees with the intuition that the production of longitudinal particles should be enhanced in the high-energy limit.  The appropriate projection operator that converts the partonic amplitude for $c\bar{c}\gamma$ production into the production of a $J/\psi$ and a photon is the following~\cite{Beneke:2001}:
\begin{equation}
{\cal M}^{LCDA}_{J/\psi,direct} = -\frac{f_{J/\psi}}{4} \frac{\mpsi}{E_{J/\psi}} \int_0^1 du \, \text{Tr}[({\cal M}_{c\bar{c}\gamma})   
\slashed{p}_{J/\psi} v\cdot \varepsilon^{*}_{J/\psi}],
\label{eq:LCDAproj}
\end{equation}
where $v=(n+\overline{n})/2$, $f_{J/\psi}$ is the decay constant of the $J/\psi$ and $\phi_{\parallel}(u)$ is the twist-2 LCDA of the $J/\psi$.  We have neglected higher-twist contributions to this projection operator.  The detailed algebraic steps indicate how the expression in Eq.~(\ref{eq:LCDAproj}) is converted into its final form:
\begin{eqnarray} \label{eq:LCDA}
{\cal M}^{LCDA}_{J/\psi,direct}
&=& \int^{1}_{0}du\frac{-eQ_{c}g_{Z}}{\bar{u}(M^{2}_{Z}-\mpsi^{2})}
(-\frac{f_{J/\psi}\mpsi}{4E_{J/\psi}})(v\cdot\varepsilon^{*}_{J/\psi}) \nonumber\\
&\times & \mathrm{Tr}\left[\slashed{\varepsilon}_{Z}(g^{c}_{V}-g^{c}_{A}\gamma_{5})
(\bar{u}\slashed{p}_{J/\psi}+\slashed{p}_{\gamma}-m_{c})\slashed{\varepsilon}^{*}_{\gamma}\slashed{p}_{J/\psi} \right] \nonumber \\
&+& \int^{1}_{0}du\frac{eQ_{c}g_{Z}}{u(M^{2}_{Z}-\mpsi^{2})}
(-\frac{f_{J/\psi}\mpsi}{4E_{J/\psi}})(v\cdot\varepsilon^{*}_{J/\psi}) \nonumber\\
&\times & \mathrm{Tr}\left[\slashed{\varepsilon}^{*}_{\gamma}(u\slashed{p}_{J/\psi}+\slashed{p}_{\gamma}+m_{c})\slashed{\varepsilon}_{Z}
(g^{c}_{V}-g^{c}_{A}\gamma_{5})\slashed{p}_{J/\psi}\right] \nonumber \\
&=& \frac{ieQ_{c}g_{Z}g^{c}_{A}f_{J/\psi}m_{J/\psi}}{M^{2}_{Z}}
\int^{1}_{0}du\left[\frac{\phi_{\parallel}(u)}{\bar{u}}+\frac{\phi_{\parallel}(u)}{u} \right]\epsilon_{\alpha\mu\nu\rho}
\varepsilon^{\alpha}_{Z}\varepsilon^{*\mu}_{\gamma}
\varepsilon^{*\nu}_{J/\psi}p^{\rho}_{\gamma} \nonumber \\
&=& \frac{ieQ_{c}g_{Z}g^{c}_{A}f_{J/\psi}m_{J/\psi}}{M^{2}_{Z}}\int^{1}_{0}du\frac{\phi_{\parallel}(u)}{u\overline{u}}\epsilon_{\alpha\mu\nu\rho}
\varepsilon^{\alpha}_{Z}\varepsilon^{*\mu}_{\gamma}
\varepsilon^{*\nu}_{J/\psi}p^{\rho}_{\gamma}.
\end{eqnarray}
In the second and fourth line, we use the relations $p_{1}= u p_{J/\psi}$ and $p_{2}= \bar{u} p_{J/\psi}$ valid in the limit $M_Z \gg \mpsi$. In the fifth line, we use the relation
\begin{equation}
(v\cdot\varepsilon^{*}_{J/\psi})\epsilon_{\alpha\mu\nu\rho}
\varepsilon^{\alpha}_{Z}\varepsilon^{*\mu}_{\gamma}
p^{\nu}_{J/\psi}p^{\rho}_{\gamma}\simeq
\frac{M_{Z}}{2}\epsilon_{\alpha\mu\nu\rho}
\varepsilon^{\alpha}_{Z}\varepsilon^{*\mu}_{\gamma}
\varepsilon^{*\nu}_{J/\psi}p^{\rho}_{\gamma}
\end{equation}
also valid in the limit $M_Z \gg \mpsi$.  Before proceeding we make a few comments on the region of validity of this result.  We have performed a light-cone expansion of all momenta appearing in the problem.  For example, $p_1$ and $p_2$ are assumed to lie along $n$, and components along $\bar{n}$ and perpendicular components are neglected.  We have also neglected higher-twist wave-functions in the projector of Eq.~(\ref{eq:LCDAproj}). Both effects are suppressed by powers of $\mpsi^2/M_Z^2$.

We have not yet used the fact that the quarks making up the $J/\psi$ are non-relativistic in the $J/\psi$ rest frame. This fact implies a connection between the decay constant $f_{J/\psi}$, the integral over $\phi_{\parallel}(u)$ and the relativistic corrections found in the previous section.  This connection was derived in detail in Refs.~\cite{Wang:2013ywc,Ma:2014}.  Converting the results of this reference into our notation, we have
\begin{eqnarray}
f_{J/\psi} &=& 2 \sqrt{\frac{N_c}{\mpsi}}  \phi_{0}(J) \left( 1-\frac{\langle v^{2} \rangle}{6}+{\cal O}(v^4)\right),\nonumber \\
\int^{1}_{0}du\frac{\phi_{\parallel}(u)}{u\overline{u}} &=& 4 \left(1+\frac{\langle v^{2} \rangle}{3}+{\cal O}(v^4)\right).
\end{eqnarray}
Only the non-relativistic limit of this expression is given in Ref.~\cite{Wang:2013ywc,Ma:2014}.  We have reproduced this limit and also have derived the ${\cal O}(v^2)$ correction following the technique of Ref.~\cite{Bodwin:1407}.  Inserting this result into Eq.~(\ref{eq:LCDA}), we reproduce both the non-relativistic limit and leading $v^2$ correction of Eq.~(\ref{eq:ampdirrel}).

The usefulness of considering the LCDA approach is that the ${\cal O}(\alpha_s)$ corrections to the direct amplitude have been calculated in Ref.~\cite{Wang:2013ywc,Ma:2014} in the non-relativistic limit, and can be included in our calculation.  The correction factor 
is given by
\begin{equation}
\label{eq:QCDcorr}
\Delta_{QCD}(\mu,\mu_0) = \frac{\alpha_s(\mu)}{4\pi}C_F \left[ (3-2\,\text{ln} \,2) \left( \text{ln}\frac{\mu^2}{\mu_0^2} -i \pi\right)
	+\text{ln}^2 2-\text{ln}\, 2-9-\frac{\pi^2}{3}\right].
\end{equation}  
The central values for the scales $\mu$ and $\mu_0$ are $\mu \sim M_z$, $\mu_0 \sim m_c$.  The logarithm comes from collinear-gluon emission.  The hard scale for this logarithm is the hard scale of the process, $\mu$, while the low scale $\mu_0$ denotes where the collinear emissions are cut off.  If desired, these logarithms could be resummed using the evolution equation satisfied by the LCDA.  
We note that the leading logarithmic correction in Eq.~(\ref{eq:QCDcorr}) gives a 17\% correction to the amplitude.  We can estimate the effect of higher-order logarithmic corrections by exponentiating this correction.  This leads to an additional 1.5\% shift beyond what has already been calculated.  Since this estimate turns out not to be large, we do not include this resummation here.

Using this next-to-leading order QCD correction, we can write down our final expression for the direct amplitude, including both relativistic and ${\cal O}(\alpha_s)$ improvements:
\begin{equation}
\label{eq:finaldir}
{\cal M}_{J/\psi,direct} = {\cal M}^{(0)}_{direct} \left[1+\frac{\langle v^{2} \rangle}{6}+\Delta_{QCD}(\mu,\mu_0)\right].
\end{equation}
We note that we have not included any mixed corrections of the form ${\cal O}(\alpha_s v^2)$ in this expression.  We will estimate later the theoretical uncertainty arising from these missing terms, as well as from other sources of error.  To summarize, this expression includes all ${\cal O}(\alpha_s)$ and ${\cal O}(v^2)$ corrections.  All terms of ${\cal O}(\mpsi^2/M_Z^2)$, ${\cal O}(\alpha_s^2)$, ${\cal O}(v^4)$ and ${\cal O}(\alpha_s v^2)$ are neglected.

\subsection{The indirect amplitude}
\label{sec:indirectamp}

We now consider the indirect contribution to the $J/\psi$ channel which arises through the diagrams in the right panel of Fig.~\ref{fig:diagrams}.  We begin by deriving the effective $Z\gamma\gamma^{*}$ vertex which mediates this process.  This coupling is loop-induced.  It was first considered for arbitrary fermions propagating in the loop in Ref.~\cite{Hagiwara:1991}.  In our notation, the amplitude for a given internal fermion $f$ is  
\begin{equation}
{\cal M}^f_{\alpha\mu\nu}(Z^{\alpha}\rightarrow \gamma^{\mu}(p_{\gamma}) \gamma^{*\nu}(p_{V}))=[-ie^{2}Q^{2}_{f}g_{Z}g^f_{A}N^f_{c}]\frac{p^{2}_{V}}{\pi^{2}}I(m_{f},\mpsi,M_Z)
\epsilon_{\alpha\mu\nu\rho}p^{\rho}_{\gamma},
\end{equation}
where $N_c^f$ denotes the number of colors for the fermion $f$. $I(m_f,\mpsi,M_Z)$ denotes the parametric integral
\begin{equation}
I(m_{f},\mpsi,M_Z)=-\int^{1}_{0} dx \int^{1-x}_{0} dy \frac{-y+y^{2}+xy}{m^{2}_{f}-y(1-y)\mpsi^{2}-xy(M^{2}_{Z}-\mpsi^{2})}.
\end{equation}
This function depends on the internal-fermion mass $m_f$, and on $\mpsi$ and $M_Z$.  We use directly this parametric integral in our numerical results.  The analytic expression for $I(m_f,\mpsi,M_Z)$ is given in Ref.~\cite{Hagiwara:1991}; we do not reproduce it here, although we have re-derived and confirmed it in several limits.  As one check, we have confirmed that our numerical result reproduces the following analytic expression in the limit of zero internal fermion mass:
\begin{equation}
I(0,\mpsi,M_Z) = \frac{1}{M_Z^2-\mpsi^2} \left[ 1+\frac{M_Z^2}{M_Z^2-\mpsi^2} \text{ln}\left( \frac{\mpsi^2}{M_Z^2}\right)\right].
\end{equation}

We note that in the limit of degenerate fermion masses within an entire generation of fermions, the amplitude becomes proportional to
\begin{equation}
\sum_{f} {\cal M}^f_{\alpha\mu\nu}(Z\rightarrow \gamma(p_{\gamma}) \gamma^{*}(p_{V})) \propto \sum_f Q^{2}_{f} g^f_{A}N^f_{c} = 0.
\end{equation}
This expression vanishes because of anomaly cancellation in the Standard Model.

To obtain the indirect amplitude for $Z \to J/\psi+\gamma$, we combine the $Z\gamma\gamma^{*}$ result with the transition amplitude for $\gamma^{*} \to J/\psi$. This transition proceeds through the following matrix element: 
\begin{equation}
i{\cal M}_{J_V}^{\mu}=-ie\langle J/\psi(\epsilon)|J_V^\mu(x=0)|0\rangle =-ieg_{J/\psi\gamma}
\epsilon^{\mu*},
\end{equation}
where $J_V$
is the electromagnetic current,
\begin{equation}
J_V^\mu(x)= \sum_q Q_q \bar q(x)\gamma^\mu q(x).
\label{eq:em-current}
\end{equation}
In Eq.~(\ref{eq:em-current}), the sum is over all quark flavors. We can solve for the magnitude of the effective coupling $g_{J/\psi\gamma}$ using the decay width of the $J/\psi$ into leptons:
\begin{equation}
\Gamma[V\to l^+l^-]=\frac{4\pi\alpha^2 g_{V\gamma}^2}{3m_V^3}.
\label{leptonic-width}
\end{equation}
In order to 
determine the phase of $g_{V\gamma}$,  we follow Ref.~\cite{Bodwin:2013gca} and note that to leading order in 
$\alpha_s$ and $v$ we have
\begin{equation}
g_{J/\psi\gamma}=-Q_c\sqrt{2N_c}\sqrt{2\mpsi}\, \phi_0(J).
\label{gvgam-phi0}
\end{equation}
This indicates that $g_{J/\psi\gamma}$ is negative\footnote{There is a small phase generated by high-order terms in the NRQCD expansion that are negligible given other theoretical uncertainties.}.  We find the following expression for the indirect amplitude:
\begin{eqnarray}
{\cal M}_{J/\psi,indirect} &=& \frac{i e^3 g_Z g_{J/\psi \gamma}}{\pi^2} \epsilon_{\alpha\mu\nu\rho}\varepsilon^{\alpha}_{Z}\varepsilon^{*\mu}_{\gamma}
\varepsilon^{*\nu}_{J/\psi} 
p^{\rho}_{\gamma} \sum_f Q_f^2 g_A^f N_c^f I(m_f,\mpsi,M_Z).
\label{eq:finalindir}
\end{eqnarray}
The sum is over all fermions in the Standard Model.  We will study the numerical impact of this contribution in a later section, but we make a few comments here.  Since the contributions are proportional to the mass splittings within a generation, we find that the first generation gives a negligible result.  The integral $I(m_f,\mpsi,M_Z)$ goes like $1/m_f^2$ for heavy fermions, and the contribution from the top quark is therefore also small.  Only the charm-quark, bottom-quark, and tau-lepton contributions are numerically important.  These contributions can be expanded in the ratio of the fermion masses over $M_Z$.  Since these ratios are small, the indirect amplitude is small for this process.  This is in contrast to the Higgs decay~\cite{Bodwin:2013gca}, for which the indirect amplitude gives a sizable contribution.  The overall QED coupling term $e^3$ contains an $e^2$ that comes from the coupling of the off-shell $\gamma^{*}$, and a factor of $e$ that comes from the on-shell photon.  We will therefore replace this factor by the following combination of running coupling constants in the $\bar{MS}$ scheme: $e^3 \to e^2(\mpsi) \times e(0)$.  We do not assign any theoretical error due to missing higher-order corrections to the indirect amplitude, since it anyway gives a small contribution to the branching ratio.

\subsection{Summary of the $J/\psi$ mode}

To form the entire amplitude for $Z \to J/\psi+\gamma$, we sum the direct and indirect amplitudes given in Eqs.~(\ref{eq:finaldir}) and~(\ref{eq:finalindir}):
\begin{equation}
{\cal M}_{Z \to J/\psi+\gamma} = {\cal M}_{J/\psi,indirect}+{\cal M}_{J/\psi,direct}.
\end{equation}
We form the partial width for this mode as
\begin{eqnarray}
\Gamma_{Z \to J/\psi+\gamma}
&=&  \frac{1}{3}\frac{1}{2M_{Z}}\frac{M^{2}_{Z}-m^{2}_{\mpsi}}{8\pi M^{2}_{Z}}
\sum_{pols} |{\cal M}_{Z \to J/\psi+\gamma}|^2 \nonumber \\
&=& \frac{1}{48 \pi M_Z} \sum_{pols} |{\cal M}_{Z \to J/\psi+\gamma}|^2.
\end{eqnarray}
On the right-hand side of the first equation, the first factor $1/3$ comes from the $Z$-boson polarization averaging, the second factor comes from the overall flux factor, and the third factor comes from the phase space, which we have expanded to leading order in $\mpsi^{2}/M^{2}_{Z}$.

The sum is over the polarizations of all three particles in the process.  We have included a $1/3$ from the $Z$-polarization averaging in this expression, and have expanded the phase-space factor to leading order in $\mpsi^2/M_Z^2$ to maintain consistency with our calculation of the amplitude.  We will discuss our numerical inputs into this partial width in a later section.  We note that the $J/\psi$ states produced are predominantly longitudinally polarized.  Transverse polarizations are suppressed by a factor of $\mpsi^2/M_Z^2$.

\section{The decay $Z\rightarrow \phi + \gamma$}
\label{sec:phi}

In this section we discuss the decay $Z \to \phi+\gamma$.  We assume that the $\phi$ meson is composed entirely of an $s\bar{s}$ pair.  In the rest frame of the $\phi$ meson the quarks are energetic and boosted along the direction of the $\phi$ momentum.  We therefore use the LCDA approach of Section~\ref{sec:LCDA} to calculate the direct amplitude for this process.  There is also an indirect contribution that we calculate similarly as the $J/\psi$ indirect amplitude of Section~(\ref{sec:indirectamp}).

\subsection{The direct amplitude}

We begin by discussing the direct amplitude.  Denoting the $\phi$ momentum as $p$ and the photon momentum as $q_1$, we expand all momenta around the light-cone directions as we did for 
the $J/\psi$ in Eq.~(\ref{eq:LCDAmom}):
\begin{eqnarray}
p^{\mu}_{\gamma} &=& \frac{M^{2}_{Z}-\mphi^2}{2M_{Z}} \bar{n}^{\mu},\;\;\; p^{\mu}_{V} = 
	\frac{M_{Z}}{2}n^{\mu}+\frac{\mphi^2}{2M_{Z}}\bar{n}^{\mu}, \\
p^{\mu}_{1} &=& u\frac{M_{Z}}{2}n^{\mu},\;\;\;
p^{\mu}_{2} = \bar{u}\frac{M_{Z}}{2}n^{\mu}.
\end{eqnarray}
We have neglected $m_s$, the strange-quark mass, in writing down the strange and anti-strange momenta $p_1$ and $p_2$.  Since $M_Z^2 \gg m_{\phi}^2$, we can further simplify these momenta by dropping the explicit $\mphi$ terms.  We next calculate the partonic direct-amplitude diagrams of Fig.~\ref{fig:diagrams}, and use a similar projector as in Eq.~(\ref{eq:LCDAproj}) to convert the partonic amplitude into one for the $\phi$-meson:
\begin{equation}
{\cal M}^{LCDA}_{\phi,direct} = -\frac{f_{\phi}}{4} \frac{\mphi}{E_{\phi}} \int_0^1 du \, \text{Tr}[({\cal M}_{s\bar{s}\gamma})   
\slashed{p}_{V} v\cdot \varepsilon^{*}_{\phi}],
\end{equation}
Here, ${\cal M}_{s\bar{s}\gamma}$ is the partonic amplitude for the production of $s\bar{s}\gamma$.  The transverse projector again gives zero to leading order in $\mphi^2/M_Z^2$, as for the $J/\psi$. A straightforward calculation gives the result
\begin{eqnarray}
\label{eq:phidir}
{\cal M}^{LCDA}_{\phi,direct} 
&=& \frac{ieQ_{s}g_{Z}g^{s}_{A}f_{\phi}\mphi}{M^{2}_{Z}}\int^{1}_{0}du\frac{\phi_{\parallel}(u)}{u\overline{u}}\epsilon_{\alpha\mu\nu\rho}
\varepsilon^{\alpha}_{Z}\varepsilon^{*\mu}_{\gamma}
\varepsilon^{*\nu}_{\phi}p^{\rho}_{\gamma}.
\end{eqnarray}
Here, $f_{\phi}$ is the $\phi$-meson decay constant, and $\phi_{\parallel}(u)$ is the twist-2 longitudinal LCDA for the $\phi$-meson.  It depends on a renormalization scale $\mu$ that we have suppressed.

At this point the calculation differs from the LCDA calculation for the $J/\psi$.  It is not possible to relate the decay constant and $\phi_{\parallel}$ to NRQCD matrix elements.  $f_{\phi}$ can simply be taken from data.  The twist-2 distribution amplitudes can be expanded in Geigenbaur polynomials~\cite{Lepage:1979zb}:
\begin{equation}
\phi_{\parallel}(u) = 6 u \bar{u} \left[ 1+\sum_{n=2} a_n^{\parallel} C_n^{3/2}(2u-1) \right].
\label{eq:geig}
\end{equation}
Here, the $C_n^{3/2}$ are Geigenbauer polynomials.  We need the $n=0$ and $n=2$ results for our calculation; they are
\begin{equation}
C^{3/2}_{0}(u) = 1, \;\;\; C^{3/2}_{2}(u) = \frac{15}{2}u^{2}-\frac{3}{2}.
\end{equation}
We note that the this distribution amplitude is normalized so that
\begin{equation}
\int_0^1 du \, \phi_{\parallel}(u) =1.
\end{equation}
The quantities $a_n^{\parallel}$ are scale-dependent.  We take their input values at $\mu=1$ GeV from Ref.~\cite{Dimou:1212}.  These values are obtained from an average of sum-rule and lattice determinations.  In our numerical results we truncate the sum of Eq.~(\ref{eq:geig}) at $n=2$.  The higher $n$ terms are not known.  Since $a_2^{\parallel}$ does not give a large contribution to the amplitude, we expect that this truncation does not lead to a large error.  Given these expressions it is straightforward to perform the integrals over $u$ in Eq.~(\ref{eq:phidir}).

In order to include the leading-logarithmic corrections from collinear gluon emission in the amplitude, we solve the evolution equation for the $a_n^{\parallel}$ to evolve them from the input scale of 1 GeV to the hard scale $\mu \sim M_Z$ of the process.  The solutions to the renormalization-group equation for the coefficients are~\cite{Lepage:1979zb}:
\begin{equation}
a^{\parallel}_n(\mu) = \left(\frac{\alpha_{s}(\mu)}{\alpha_{s}(\mu_0)}\right)^{\frac{\gamma^{\parallel}_{n}}{2\beta_{0}}} a^{\parallel}_{n}(\mu_0),
\end{equation}
where
\begin{equation}
\gamma^{\parallel}_{n}=8C_{F}\left(\sum^{n+1}_{k=1}\frac{1}{k}-\frac{3}{4}-\frac{1}{2(n+1)(n+2)}\right).
\end{equation}
We will use this renormalization-group improved expression in our numerical results.

\subsection{The indirect amplitude}

The calculation of the indirect amplitude for $Z \to \phi+\gamma$ proceeds identically to the calculation for $Z \to J/\psi+\gamma$ presented in Section~\ref{sec:indirectamp}.  We simply take over the result from Eq.~(\ref{eq:finalindir}), replacing all $J/\psi$-dependent quantities with their $\phi$ analogues.  We obtain
\begin{eqnarray}
{\cal M}_{\phi,indirect} &=& \frac{i e^3 g_Z \mphi f^{\phi}_{V}Q_{s}}{\pi^2} \epsilon_{\alpha\mu\nu\rho}\varepsilon^{\alpha}_{Z}
\varepsilon^{*\mu}_{\gamma}\varepsilon^{*\nu}_{J/\psi}
p^{\rho}_{\gamma} \sum_f Q_f^2 g_A^f N_c^f I(m_f,\mphi,M_Z),
\label{eq:phiindir}
\end{eqnarray}
where we have used 
\begin{equation}
\langle\phi |J^{\mu}_{V}(0)|0\rangle=f_{\phi}\mphi\varepsilon^{\mu}_{\phi}.
\end{equation}
There is again no contribution in the limit of degenerate fermion masses propagating inside the loop.  The numerically-relevant contributions come from the tau lepton, charm quark and bottom quark.

\subsection{Summary for $Z \to \phi+\gamma$}

The final expression for $Z \to \phi+\gamma$ amplitude comes from summing the direct and indirect expressions of Eqs.~(\ref{eq:phidir}) and~(\ref{eq:phiindir}):
\begin{equation}
{\cal M}_{Z \to \phi+\gamma} = {\cal M}_{\phi,indirect}+{\cal M}_{\phi,direct}.
\end{equation}
We form the partial width for this mode as
\begin{equation}
\Gamma_{Z \to \phi+\gamma} = \frac{1}{48 \pi M_Z}  \sum_{pols} |{\cal M}_{Z \to \phi+\gamma}|^2.
\end{equation}
This expression is valid to leading order in $\mphi^2/M_Z^2$, and includes the leading-logarithmic QCD corrections coming from the evolution of the $\phi$-meson twist-2 LCDA.

\section{Numerical results}
\label{sec:numerics}

We discuss in this section our numerical results for both the central values and theoretical errors for the $Z \to J/\psi+\gamma$, $Z \to \Upsilon(1S)+\gamma$, and $Z \to \phi+\gamma$ decays.  We begin with the $J/\psi$ process.  For the direct amplitude, we use the following values for the parameters which enter the prediction:
\begin{equation}
\alpha(0) = 1/137.036, \;\;\; \phi_0(J) = 0.270\pm 0.020 \, \text{GeV}^{3/2},\;\;\; \langle v^2 \rangle = 0.201\pm 0.064.
\end{equation}
The values for the $J/\psi$ wave-function at the origin, $ \phi_0(J)$, and the matrix element $\langle v^2 \rangle$ are taken from Ref.~\cite{Bodwin:2007fz,Bodwin:1407}.  We have made the replacement $e \to \sqrt{4\pi \alpha(0)}$ in Eq.~(\ref{eq:nonrel}), as appropriate for an on-shell photon.  When evaluating the QCD correction presented in Eq.~(\ref{eq:QCDcorr}), we choose the central scales $\mu=M_Z$ and $\mu_0 = m_c$.  For numerical consistency with the results of Ref.~\cite{Bodwin:1407} which studies Higgs decays, we convert the $\overline{MS}$ charm mass to the pole mass at one-loop order.  We use the $\overline{MS}$ mass and error coming from the Particle Data Group (PDG)~\cite{Agashe:2014kda} as input, and to perform the conversion to the pole mass we use {\tt RunDec}~\cite{Chetyrkin:2000yt}.  We do the same for the bottom-quark mass, which is needed in the evaluation of the indirect amplitude.  We find the following result for the pole masses:
\begin{equation}
m_c = 1.485 \pm 0.026 \, \text{GeV},\;\;\; m_b = 4.579 \pm 0.032 \, \text{GeV}.
\end{equation}

For the indirect amplitude, we further need to specify $g_{J/\psi\gamma}$ and the masses of the fermions that propagate in the loop.  The charm and bottom masses are given above.  We use the PDG value for the tau-lepton mass.  All other contributions are numerically negligible.  For the coupling $g_{J/\psi\gamma}$, we use the result of Eq.~(\ref{leptonic-width}) and take $\Gamma[J/\psi\to l^+l^-]$ and its experimental error from the PDG.  We obtain
\begin{equation}
g_{J/\psi\gamma} = -0.832 \pm 0.010\, \text{GeV}^2.
\end{equation}
For all other parameters that appear in the $J/\psi$ amplitude, we use the values from the PDG.

In order to estimate the theoretical errors on the $J/\psi$ branching ratio, we consider the following sources of uncertainty.
\begin{itemize}

\item We study parametric uncertainties arising from $\phi_0(J)$, $\langle v^2 \rangle$, $g_{J/\psi\gamma}$, $m_c$, and $m_b$. 

\item We estimate the uncertainty coming from uncalculated ${\cal O}(\alpha_s^2)$ corrections by varying the scale $\mu$ in the direct amplitude in the range $\mu \in [M_Z/2, 2 M_Z]$.

\item We estimate the uncertainty from higher-order terms in the NRQCD expansion by assigning a relative uncertainty of $\langle v^2 \rangle^2$ to such corrections.

\item We estimate the uncertainty on mixed ${\cal O}(\alpha_s v^2)$ corrections by assigning a relative uncertainty of $ \alpha_s /(4\pi) \times \langle v^2 \rangle$ to these corrections.

\end{itemize}
We will see that the indirect amplitude gives a small contribution to the branching ratio, justifying our neglect of a theoretical error on this term.  All of these sources of uncertainty are added in quadrature separately for both the direct and indirect amplitudes to produce an uncertainty on each contribution.   Both the direct and indirect amplitudes are then allowed to vary independently within their allowed errors, and the envelope of these deviations is then taken to produce a final error on the branching ratio prediction.  

Using the prescriptions above, and taking the total width of the $Z$-boson from the PDG, we arrive at the following prediction for the J/$\psi$ branching ratio in the Standard Model:
\begin{equation}
B_{SM}(Z \to J/\psi+\gamma) = (9.96 \pm 1.86) \times 10^{-8}.
\end{equation}
We make a few comments on this result.  If the indirect amplitude were set to zero, the central value of the branching ratio would instead be $1.00 \times 10^{-7}$.  The indirect amplitude interferes destructively with the direct amplitude, but leads to only a 0.4\% decrease in the result.  The largest correction to the leading non-relativistic result of Eq.~(\ref{eq:nonrel}) comes from the $v^2$ correction of Eq.~(\ref{eq:ampdirrel}).  If this were turned off, the branching ratio would decrease by 6\%.  The relative error on the branching ratio is 18.7\%.  In order of importance, the three largest contributions to the error budget are the parametric uncertainty on $\phi_0(J)$, our estimate of the missing $v^4$ corrections, and the scale variation in the direct amplitude.  If the error on $\phi_0(J)$ is removed, the relative uncertainty decreases to only 10.3\%.  This parametric uncertainty dominates the error budget.  If all three sources of uncertainty are turned off, the relative uncertainty becomes only 2.1\%.

We study next the decay $Z \to \Upsilon(1S)+\gamma$, which is very similar to the $J/\psi$ mode just considered.  The primary difference in this case is that the quarkonium mass effects that we have neglected go like $m_{\Upsilon}^2/M_Z^2$, which reaches the percent level.  Since this is still a small correction, we continue to neglect such effects.  We use the following values for the parameters which enter the direct amplitude:
\begin{equation}
\alpha(m_{\Upsilon}) = 1/131.87, \;\;\; \phi_0(\Upsilon) = 0.715\pm 0.024 \, \text{GeV}^{3/2},\;\;\; \langle v^2 \rangle = -0.00920\pm 0.00348.
\end{equation}
The values for wave-function at the origin and the matrix element $\langle v^2 \rangle$ are taken from Ref.~\cite{Bodwin:2007fz,Bodwin:1407}.  For the indirect amplitude, we further need to specify $g_{\Upsilon\gamma}$.  We use the result of Eq.~(\ref{leptonic-width}) and take $\Gamma[\Upsilon\to l^+l^-]$ and its experimental error from the PDG.  We obtain
\begin{equation}
g_{\Upsilon\gamma} = 2.212 \pm 0.015\, \text{GeV}^2.
\end{equation}
We consider the same sources of uncertainty  as for the $J/\psi$.  We arrive at the following prediction:
\begin{equation}
B_{SM}(Z \to \Upsilon(1S)+\gamma) = (4.93 \pm 0.51) \times 10^{-7}.
\end{equation}
The indirect amplitude again has a sub-1\% effect on this branching ratio.  The largest sources of uncertainty are the parametric error on $\phi_{0}(\Upsilon)$ and the scale variation.

We now consider the decay $Z \to \phi+\gamma$.  The branching ratio for this process additionally depends on the decay constant $f_{\phi}$ and the coefficient $a_2^{\parallel}$ that appears in the twist-2 LCDA.  We take these quantities from Ref.~\cite{Dimou:1212}:
\begin{equation}
f_{\phi} = 0.235 \pm 0.005\, \text{GeV},\;\;\; a_2^{\parallel}(1 \, \text{GeV})=0.23 \pm 0.08.
\end{equation}
For our error estimate we consider parametric uncertainties coming from $f_{\phi}$, $a_2^{\parallel}$, $m_c$, and $m_b$. We estimate missing higher-order corrections in the direct amplitude by taking $\mu \in [M_Z/2, 2 M_Z]$.  We find the result:
\begin{equation}
B_{SM}(Z \to \phi+\gamma) = (1.17 \pm 0.08) \times 10^{-8}.
\end{equation}
The indirect amplitude has a small effect on this branching ratio; setting it to zero leads to a 1\% increase in the branching ratio.  However, the evolution of $a_2^{\parallel}$ from the input scale of 1 GeV to $\mu=M_Z$ has a large effect on the rate.  Without this effect, the branching ratio would be $1.51 \times 10^{-8}$, almost 30\% larger.  The dominant sources of uncertainty are the parametric errors on $f_{\phi}$ and $a_2^{\parallel}$.  If these were removed, the remaining estimated error would drop to the 
1.3\%.

\section{Conclusions}
\label{sec:conc}

In this manuscript we have studied the rare decays $Z \to J/\psi+\gamma$, $Z \to \Upsilon+\gamma$ and $Z\to \phi+\gamma$.  Our motivation in considering these processes is that they serve as benchmark processes for similar rare decays of the Higgs boson to mesons that may reveal whether the Yukawa structure in Nature is indeed that predicted  by the Standard Model.  We have performed a detailed study of all contributions which lead to these rare $Z$-boson decays, including both the direct and indirect amplitudes.  For the heavy quarkonium decays we utilized the NRQCD framework, and included the leading relativistic and ${\cal O}(\alpha_s)$ corrections.  For the $\phi$ decay we used the LCDA approach and included the leading-logarithmic QCD corrections.  In both cases we carefully considered all sources of parametric and theoretical uncertainties.  The dominant uncertainties for both processes are parametric in origin: for the $J/\psi$ and $\Upsilon$ modes the largest error is from knowledge of the wave-functions, while for the $\phi$-meson the two largest sources are the decay constant $f_{\phi}$ and the LCDA itself.  If necessary, it may be possible to reduce these in the future with a combination of improved experimental data and lattice calculations.

Although small, there is a probability that the $J/\psi$ and $\Upsilon$ decays will be observed at Run II of the LHC.  The final state is clean, consisting of two leptons and a photon that reconstruct to the $Z$ mass peak if a leptonic decay of the quarkonium is demanded.  Although this further reduces the branching ratio, the inclusive $Z$ production cross section at even the 8 TeV LHC is 34 nb~\cite{Chatrchyan:2014mua}.  Almost $10^9$ $Z$-bosons were produced at the 8 TeV LHC run, and branching ratios of $10^{-7}$ should soon be accessible.  The observation of the $\phi$ decay would require a new trigger, since the $\phi$ does not have an appreciable leptonic decay.  Given the importance of this mode for the study of Higgs boson properties, we believe that the development of this trigger it should be pursued by the LHC collaborations.  We look forward to these searches being performed in the coming run of the LHC.

\smallskip
\noindent
{\bf  Acknowledgements}

\noindent
We thank K.~Nikolopoulos for helpful discussions and for encouraging us to pursue this calculation.  We also thank A.~Chisholm, S.~Stoynev, and M.~Velasco for useful discussions.  This work is supported in part by the DOE contract DE-FG02-91ER40684 and the grant DE-AC02-06CH11357.

\end{document}